# Predicting Mesoscopic Larmor Frequency Shifts in Ex Vivo Porcine Optic Nerve

*Anders Dyhr Sandgaard[1,2], André Pampel[2], Roland Müller[2], Niklas Wallstein[2,3], Toralf Mildner[2], Carsten Jäger[4,5], Markus Morawski[5], Aage Kristian Olsen Alstrup[6,7], Harald E. Möller[2,8], Sune Nørhøj Jespersen[1,9]*

[1]*Center of Functionally Integrative Neuroscience, Department of Clinical Medicine, Aarhus University, Aarhus, Denmark,*

[2]*NMR Methods & Development Group, Max Planck Institute for Human Cognitive and Brain Sciences, Leipzig, Germany*

[3]*Aix Marseille Univ, CNRS, CRMBM, Marseille, France*

[4]*Department of Neurophysics, Max Planck Institute for Human Cognitive and Brain Sciences, Leipzig, Germany,*

[5] *Paul Flechsig Institute - Center of Neuropathology and Brain Research, Medical Faculty, University of Leipzig, Leipzig, Germany,*

[6]*Department of Clinical Medicine, Aarhus University, Aarhus, Denmark*

[7]*Department of Nuclear Medicine, Aarhus University Hospital, Aarhus, Denmark.*

[8]*Felix Bloch Institute for Solid State Physics, Leipzig University, Leipzig, Germany*

[9]*Department of Physics and Astronomy, Aarhus University, Aarhus, Denmark*

**Corresponding Author**: Anders Dyhr Sandgaard.

**Postal address**: CFIN, Department of Clinical Medicine, Universitetsbyen 3, Building 1710, 8000 Aarhus C, Denmark.

**Mail**: anders@cfin.au.dk

## Keywords

1) Magnetic susceptibility, 2) Larmor frequency, 3) Modelling, 4) QSM, 5) µQSM

## Conflict of interest

The authors declare no conflict of interest.

## 1 | Abstract

**Purpose**

Larmor frequency shifts in white matter (WM) vary with fiber orientation due to anisotropic microstructure. Since clinical voxels are significantly larger than these microscopic frequency variations, the measured signal represents a bulk average of local shifts. Accurate estimation of magnetic susceptibility therefore requires accounting for these underlying frequency distributions that exist below the imaging resolution.

**Methods**

We evaluated whether Microstructure-informed Quantitative Susceptibility Mapping (µQSM) can predict orientation-dependent sub-voxel frequency shifts from orientationally dispersed hollow cylinders and spherical inclusions. Diffusion-weighted and multi-gradient-echo images were acquired from ex vivo pig optic nerves at multiple orientations relative to the main magnetic field using a 3T Siemens Connectom scanner. We also analyzed de-ironed optic nerves to try and separate the effects of myelin and iron on susceptibility.

**Results**

The estimated sub-voxel frequency shifts closely matched µQSM predictions, consistent with mesoscopic field perturbations generated by uniformly magnetized axons. De-ironing had minimal effect on the frequency shifts, indicating negligible iron contribution.

**Conclusion**

µQSM accurately reproduces the orientation dependence of Larmor frequency shifts in optic nerve WM, providing new insight into their microstructural origin and supporting improved estimation of tissue magnetic susceptibility in Quantitative Susceptibility Mapping.

## 2 | Introduction

Magnetic resonance imaging (MRI) is a powerful imaging technique capable of probing the magnetic properties of biological tissue[1]. Although its spatial resolution is typically in the millimeter range, MRI captures voxel-averaged signatures of the underlying magnetic and microstructural features. This makes it highly effective for detecting tissue alterations that are central to understanding neurodegenerative processes[2,3] such as axonal demyelination or degradation, iron accumulation or tissue inflammation.

To clarify how macroscopically averaged MRI signals relate to microscopic tissue properties, biophysical modeling frameworks have been developed to link measurable MRI contrasts with the underlying magnetic tissue architecture[4–8]. One such framework is Quantitative Susceptibility Mapping (QSM)[9], which aims to estimate the mean bulk tissue magnetic susceptibility in each voxel. This bulk susceptibility describes how strongly the tissue becomes magnetized and represents an average over the microscopic susceptibility of all cellular and subcellular components within the voxel. As a result, susceptibility varies with the local abundance of tissue constituents such as myelin and iron, which is believed to strongly influence magnetization[10–13]. In traditional relaxation-based MRI contrasts, discriminating between macroscopic changes such as hemorrhage or calcification[14] can be hard, as they may lead to similar signal relaxation. Magnetic susceptibility provides a wider dynamic range as hemorrhage has a positive paramagnetic susceptibility, and calcifications a negative diamagnetic susceptibility. Moreover, susceptibility contrast correlates well with histological markers of myelin and iron[15].

The QSM model aims to describe how magnetized tissue produces a spatially varying microscopic field $\Delta \boldsymbol{B}(\boldsymbol{r})$ that perturbs the uniform scanner field $\mathbf{B_0} = B_0 \hat{\mathbf{z}}$. This perturbation induces a phase shift in the complex MRI signal that reflects the average Larmor frequency shift $f_{\text{MRI}}(\mathbf{R})$ inside each voxel centered at a position $\mathbf{R}$. By estimating and subsequently inverting these frequency shifts to estimate bulk susceptibility $\overline{\chi}(\mathbf{R})$, QSM provides a means for characterizing tissue composition and is now widely used to study iron accumulation, demyelination, and other neurological conditions.

However, when relating the measured frequency shift $f_{\text{MRI}}(\mathbf{R})$ to bulk susceptibility $\overline{\chi}(\mathbf{R})$, QSM assumes the average self-induced magnetic field produced by tissue *inside* the voxel at $\mathbf{R}$ is zero[16]. As microscopic cells also generate magnetic fields inside the voxel, and because these fields arise from microscopic structures, they vary on the microscopic scale. When the magnetized structures are anisotropic, for example myelinated axons, the self-induced field is non-zero and anisotropic[16]. Consequently, the frequency shift depends on the orientation between the axonal axis $\hat{\boldsymbol{n}}$ and the scanner field axis, which in turn produces an orientation dependent bias in susceptibility estimation using QSM[17]. This makes QSM valid in principle only when the magnetization is weak and uniformly distributed in every voxel, which essentially treats tissue in each voxel as a fluid with homogeneous susceptibility. Hence, while QSM contrast is comparable to histology images, its values are only semi-quantitative in cerebral white matter (WM)[18] and nearby tissue[19].

A clear demonstration of this microstructure-related frequency shift was provided by Wharton and Bowtell (W&B)[18], who measured the frequency shift inside and surrounding a fresh porcine optic nerve. The experimental design was inspired by a similar experiment carried out by Luo et al.[20] who used fresh and fixated rat optic nerve. By placing the nerve in an MR-visible medium (such as phosphate-buffered saline (PBS) or agarose), the frequency pattern surrounding the nerve could be measured. W&B showed that the frequency shift measured outside the nerve can be used to isolate the macroscopic shift $f_{\text{QSM}}(\mathbf{R})$, as modelled by QSM, thereby separating bulk susceptibility effects from the measured Larmor frequency shift $f_{\text{MRI}}(\mathbf{R})$. By disentangling such macroscopic field effects from $f_{\text{MRI}}(\mathbf{R})$, they revealed a substantial orientation-dependent residual frequency shift $\delta f$ due to WM anisotropic *microstructure*, while anisotropic *susceptibility*[17] (relating to the difference in magnetic susceptibility perpendicular and parallel to each myelin lipid layers) accounted for only a small fraction of the total effect. Their work highlights that models aiming to refine QSM must account for field-contributions due to WM *structural anisotropy*, since they represent a major source of orientation dependent contrast, and that susceptibility anisotropy may be of less importance.

Microstructure-informed Quantitative Susceptibility Mapping[21,22], µQSM, was introduced to incorporate microstructural field effects directly into QSM. Instead of treating each voxel as a homogeneously magnetized source, µQSM introduces an analytical model solution for the "sub-voxel" mesoscopic frequency shift $f^{\text{Meso}}(\mathbf{R})$ generated by anisotropic axonal microstructure and anisotropically distributed spherical inclusions (cf. Figure 1). Similar frameworks that have incorporated $f^{\text{Meso}}$ have been presented previously, e.g. the General Lorentzian Tensor Approach[4,5]. µQSM can be seen as an extension of such models, as it accounts both for orientation dispersion, and arbitrary volume fraction of multiple susceptibility sources. The orientation dependence of $f^{\text{Meso}}$ is described entirely by the fiber orientation distribution function[23–25] (fODF), which captures axonal orientation and dispersion within each voxel. Once the fODF is estimated from diffusion-weighted MRI[23,26], only the mean magnetic susceptibility of the individual sources still remains to be determined from $f_{\text{MRI}}(\mathbf{R})$ in each voxel. Hence, estimating susceptibility with µQSM constitutes a linear inverse problem similar to QSM. As a result, µQSM has the potential to improve susceptibility estimates in regions where microstructural effects are non-zero[21].

The physical basis of µQSM has been supported by Monte-Carlo simulations in realistic axonal substrates[27]. These simulations modelled water diffusion and microscopic field perturbations in 3D WM geometries derived from electron microscopy[28,29]. There it was demonstrated that the fODF from diffusion MRI can accurately capture axonal orientation dispersion and is sufficient to predict the orientation dependence of axon induced mesoscopic frequency shifts[7,21]. Across a broad range of geometries and imaging conditions, µQSM successfully predicted the Monte-Carlo simulated microstructure-induced frequency shifts. Although these simulations provide strong support for µQSM, direct experimental validation would further corroborate that the model predicts both the amplitude and orientation dependence of the microstructure-induced frequency shifts in real tissue.

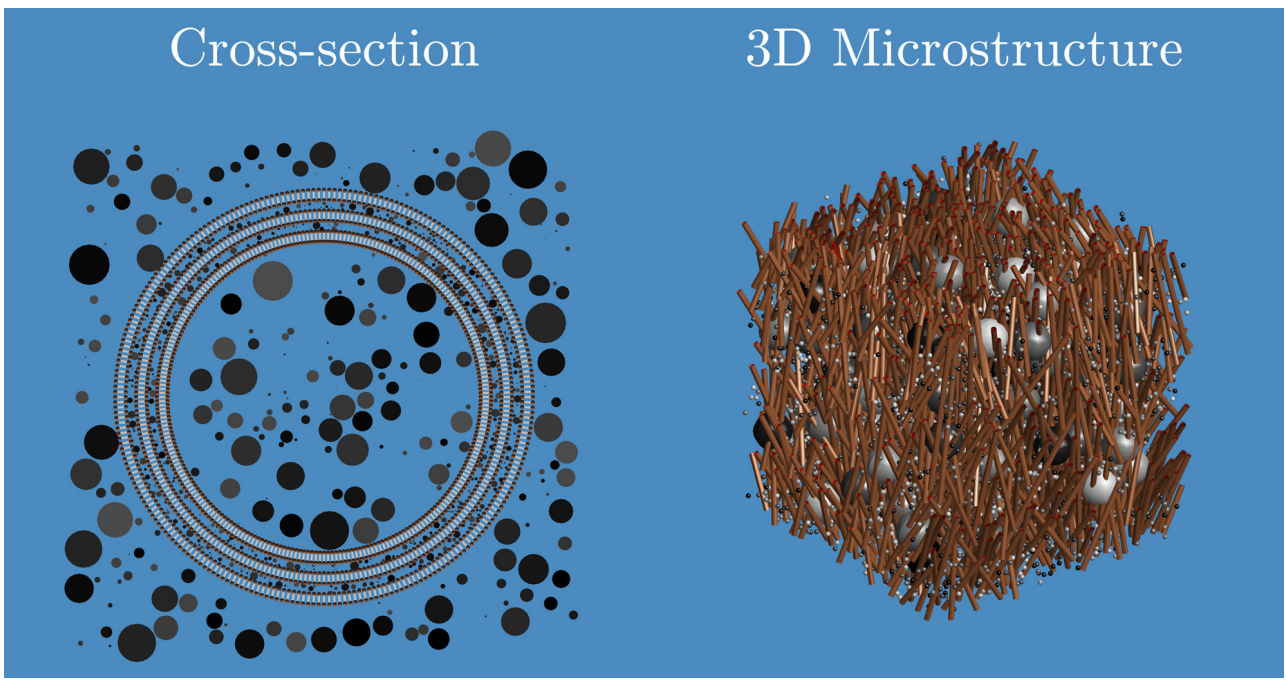

Figure 1 – Magnetic microstructure model from different perspectives: Left image conceptualizes the magnetic microstructure of a single axon with susceptibility $\chi^C$. Spherical inclusions with isotropic susceptibility $\chi^S$ can be randomly positioned inside and outside cylinders and could mimic iron complexes or other sub-cellular structures, which are depicted by differently colored spheres. However, no assumptions are made regarding their position, meaning they can be outside, inside, in-between or everywhere at once[30]. Right image shows the entire 3D magnetic microstructure to demonstrate orientation dispersion and other spherical inclusions e.g., neuroglia. Proportions are exaggerated for illustrative purposes.

*Objectives of this study*

The experimental paradigm introduced by W&B[18] offers an ideal setup for validating μQSM, as their approach measures the frequency pattern surrounding WM - enabling isolation of the residual frequency shift that cannot be explained by bulk susceptibility. We postulate: The residual Larmor frequency shift $\delta f$ reflects the mesoscopic frequency shift $f^{\text{Meso}}$ generated by axonal microstructure with scalar susceptibility and can be predicted by μQSM. To address this statement, we perform the following two main analyses:

**Objective O1) Predicting the residual Larmor frequency shift in WM.**

We reproduce the experiment of W&B[18] to confirm the presence of a residual frequency shift $\delta f$ in WM, which cannot be described by QSM. The tissue-averaged bulk susceptibility $\bar{\chi}^{\text{WM}}$ is estimated without microstructural bias by analyzing the frequency shift outside the tissue, both before and after de-ironing to also evaluate the influence of iron. The residual $\delta f$ inside the tissue is then compared with the mesoscopic shift $f^{\text{Meso}}$, which is estimated by μQSM using the fitted susceptibility $\bar{\chi}^{\text{WM}}$ and the independently estimated fODF from a dMRI experiment.

**Objective O2) Susceptibility estimation from whole sample fitting using QSM and μQSM.**

The measured frequency shift $f_{\text{MRI}}(\mathbf{R})$ is fitted across the entire sample (i.e. inside and outside the optic nerve) with μQSM, which accounts for sub-voxel frequency shift, and QSM which does not. The susceptibility estimates from QSM and μQSM are compared with the "ground truth" susceptibility $\bar{\chi}^{\text{WM}}$ established in **O1)**.

The frequency shift $f_{\text{MRI}}$ is measured at multiple sample orientations to the external field $\mathbf{B_0}$, and all are fitted at once to estimate the magnetic susceptibility in O1) and O2).

# 3 | Methods

All animal experiments were preapproved by the competent institutional and national authorities and carried out according to European Directive 2010/63.

In the spirit of W&B[18], we harvested porcine optic nerve immediately after euthanasia. The pigs (Danish Landrace and Yorkshire sows, approximately 40 kg and 100 days old) were acclimatized for at least one week before being used in a preliminary PET scan experiment under ketamine-midazolam-propofol anesthesia. The PET tracers were all given in tracer concentrations and used for the development of new diagnostic tools. The pigs were euthanized with an overdose of pentobarbital, after which the pig's skull was cut with a small-toothed saw so that the two optic nerves could be exposed and dissected free with scissors, forceps and scalpels. Pigs from PET scan experiments were used based on the 3R principle of reducing the number of animals used in research[31], and that such PET scan experiments in particular are not expected to affect the optic nerves of the pigs. The nerves were selected due to their predominantly myelinated and coherently oriented axons. Here we present the results of the final optimized iteration, but results from a preceding experiment are also reported in supplementary material for transparency.

**Sample Preparation**

A pair of optic nerves were harvested from the same porcine directly after euthanasia. Extraction took approximately 30 minutes. The nerves were then immersed in phosphate-buffered saline (PBS) containing 3% paraformaldehyde (PFA) and 1% glutaraldehyde (GA). PFA provided rapid fixation throughout the tissue, while GA ensured long-term structural stability[32–34]. The nerves remained refrigerated (4°C) in fixative for two months (fixative changed after 1 and 2 weeks), after which they were transferred to PBS to wash out residual fixative[34]. Optic nerve has a collagen-rich dural sheath surrounding the nerve[35], resembling the epineurium surrounding peripheral nerves, which we removed. While W&B used fresh optic nerves, we chose fixed optic nerves since MRI took around two days to acquire per nerve.

One of the two nerves underwent a de-ironing procedure as described in Brammerloh et al.[36]: We incubated the nerve in the de-ironing solution that was refreshed regularly (every third day) for 12 days until the solution remained clear, indicating no further iron removal. Each nerve was then placed inside spherical container filled with PBS (cf. Figure 2). The inner and outer sphere diameter were 24 and 28 mm, respectively. PBS provided an MRI-visible media for detecting frequency shifts outside the tissue. The nerves were suspended using synthetic surgical suture. To reduce bubbles on the tissue surface, the sphere was shaken and underwent vacuum degassing. To reduce sphere surface signal artifacts, the interior of the sphere was coated with Fomblin® (Solvay Solexis, Bollate, Italy) prior to tissue placement and PBS filling.

From each nerve, 2-mm thick coronal sections were cut off from one end for transmission electron microscopy (TEM), using a sharp blade. The samples were contrasted in 2% osmium tetroxide in cacodylate buffer at room temperature for 1 h, rinsed in cacodylate buffer, dehydrated by a graded series of acetone, contrasted in 1% uranyl acetate, and embedded in Durcupan™ araldite casting resin M (Fluka; Sigma-Aldrich, Buchs, Switzerland). For structural orientation, semithin sections were cut at 1 μm thickness, stained with Toluidine Blue (Merck, Darmstadt, Germany) and scanned with an AxioScan Z.1 (40×, NA 0.95; Carl Zeiss AG, Oberkochen, Germany). Ultrathin sections were cut on a Reichert Ultramicrotome II and imaged with a LEO EM 912 Omega TEM (Carl Zeiss AG) at 80 kV. Digital micrographs were obtained with a dual-speed 2 K-on-axis CCD camera-based YAG scintillator and the software ImageSP (version 1.2.13.17; TRS-Tröndle, Moorenweis, Germany). Figure 3 shows an example TEM image, demonstrating intact axons with well-preserved myelin sheaths.

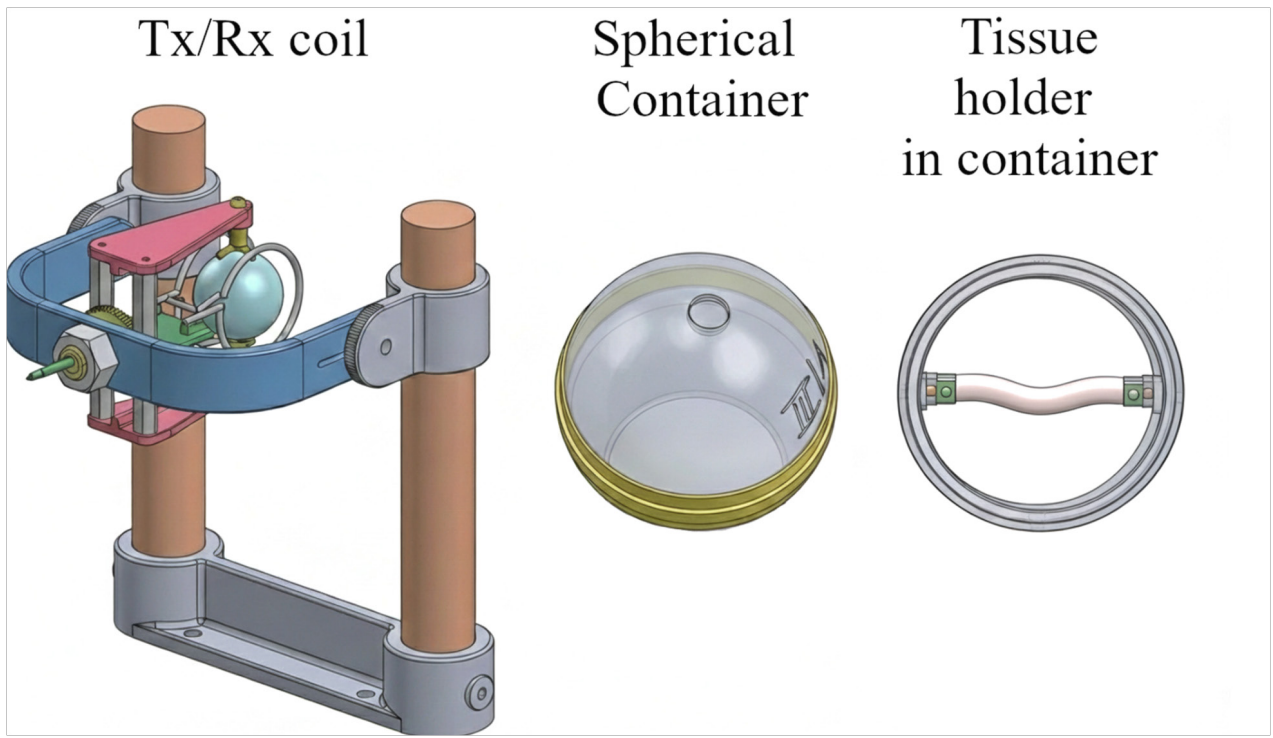

Figure 2 – Experimental setup: A spherical plastic container is held at the center of a custom-made transceive Helmholtz coil[37]. The blue plastic handle can be turned to achieve 2D coil rotations. The 3D-printed sphere contains a piece of pig

optic nerve (represented by the tube held at its end to the right) held in place at its two ends by wires. The sphere was filled with PBS through the small hole in the top and taped to seal off.

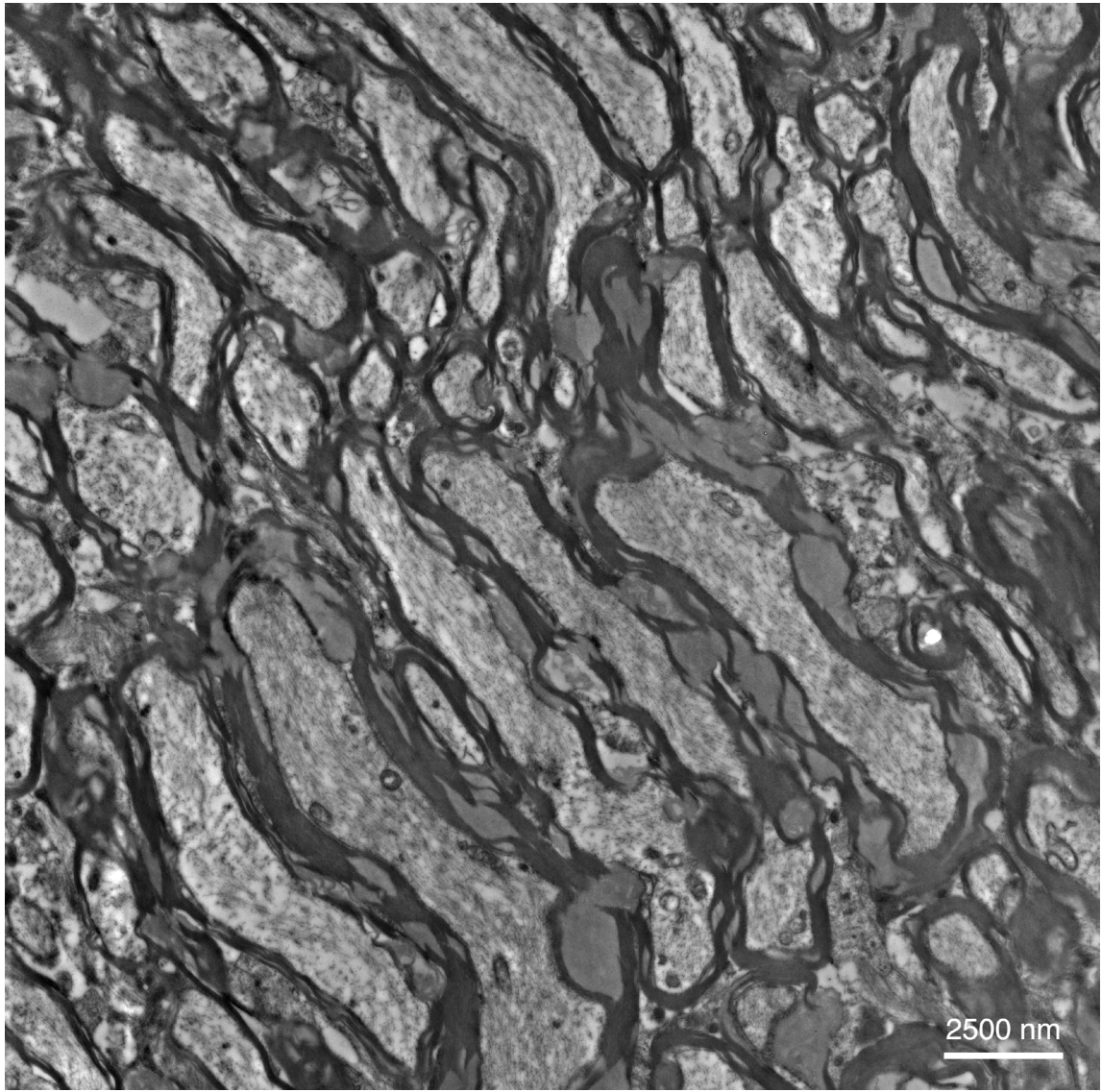

Figure 3 - Example of a TEM image (80 kV, 4000× magnification) of a section (23 μm × 23 μm) of the optic nerve (without iron removal), showing intact axons with well-preserved myelin sheaths. Because the axons were cut at a slight angle, the cross-sections appear elongated along the direction of the cut.

**Ex vivo MRI**

All image reconstruction, processing and analysis was done in MATLAB (R2025. The MathWorks, Natick, MA, USA).

Ex vivo imaging was performed at 3T on a MAGNETOM Skyra Connectom A (Siemens Healthineers, Forchheim, Germany) with a custom-built, rotatable transmit-receive coil (Figure 2)[37]. The setup enabled high-resolution imaging at 3T, and the rotation system ensured that the RF-coil was positioned in the magnet's isocenter the same way with respect to the sample for every orientation.

*Multi gradient echo acquisition*

We acquired 3D multi-gradient-echo (MGE) data with a monopolar readout gradient at approximately 250 µm isotropic nominal resolution along 12 different sample orientations in the sagittal plane achieved through coil rotations from 0° to 90° in increments of 7.5° and in random order. MGE echo times ranged from 7.05 ms to 66.85 ms in steps of 11.96 ms. MGE acquisition took around 25 minutes per orientation. The complex MGE data were MPPCA-denoised[38] and subsequently Gibbs-unrung[39]. The MRI-measured Larmor frequency shift was estimated from the last three echoes (42.93–66.85 ms) to reduce myelin-water contributions. The voxel-wise shift at $\mathbf{R}$ is denoted $f_{\mathrm{MRI}}(\mathbf{R}; \hat{\mathbf{B}})$; stacking over all $N$ voxels in the FOV gives the $N \times 1$ vector $\vec{f}_{\mathrm{MRI}}(\hat{\mathbf{B}}) = \mathrm{vec}(f_{MRI}(\mathbf{R}; \hat{\mathbf{B}}))$. Position $\mathbf{R}$ is written explicitly when needed and otherwise kept implicit. Next, $\vec{f}_{\mathrm{MRI}}$ was phase-unwrapped using SEGUE[40] and background fields were removed using LBV (with depth and peel equal to 2 and tolerance $10^{-3}$. We also tested SHARP[41], VSHARP and PDF[42], and found consistent results). The frequency shifts $\vec{f}_{\mathrm{MRI}}$ were all referenced to PBS outside the tissue for each sample orientation to minimize dependence of microstructure. We also estimated the effective transverse relaxation $\vec{R}_2$ for each orientation, which we used to segment the nerve and investigate fixation and de-ironing effects.

After processing, we co-registered the data using a rigid transformation for each sample orientation to the images of the first acquired orientation. We define $\hat{\mathbf{B}}$ as the direction of the external magnetic field after co-registration, which was determined from the rotation matrix of the rigid transformation. Registration was performed using segmented masks of the spherical container for each orientation, relying on the same fixed waypoint features, such as the filling port and tissue holders, present in all orientations. The masks, which were used for registration only, were mildly smoothed with a Gaussian filter to achieve better co-registration. The rigid transformation matrix was then applied to all images acquired for each $\theta$ using cubic interpolation. Binary images, such as masks, were interpolated using only the nearest neighbor value.

*Diffusion MRI acquisition*

Diffusion-weighted data was acquired at 400 μm isotropic resolution using a highly segmented multi-slice acquisition[43]. dMRI acquisition took around 24 hours for each nerve. Diffusion encoding was performed with b=10 ms/μm² across 60 orientations[44] with an echo time of 60 ms and a diffusion time of approximately 30 ms. Magnitude dMRI data were MPPCA-denoised[38] and subsequently Gibbs-unrung[39]. We estimated the Fiber Orientation Distribution function (fODF) using Fiber Ball Imaging[23] (FBI), which is similar to the Standard Model of Diffusion in WM[26] (SM), assuming only intra-axonal water surviving the strong diffusion weighting. From the fODF, we could estimate the scatter matrix[7,45] **T**, which depends on the second moment of the fODF and describes both the main orientation of axons in voxel **R** and their orientation dispersion. We quantified the amount of dispersion by a dispersion angle $\theta_{p2}$ derived from the second moment's rotation-invariant parameter $p_2$ of the fODF[46]. For later visualization, we denote by $t$ the average major eigenvector of **T(R)** in the tissue, and $\angle(t, \widehat{B})$ is the tissue-averaged angle between the axons and external field. The estimated parameters from dMRI were also co-registered to the MGE images of the first acquired orientation. We used an affine transformation to co-register dMRI to MGE data, as differences in acquisition scheme may lead to different image distortions between the two image protocols. The transformation was then applied to each element of **T(R)** using cubic interpolation. Since each tensor element value was calculated in the original dMRI image basis, we also need to rotate the tensors themselves. This was done by extracting the pure rotation matrix from the voxel-wise affine transformation using a polar decomposition, implemented via singular value decomposition (SVD). For convenience, we write $\vec{T}(\widehat{B}) = \text{vec}(\widehat{B}^T T(R)\widehat{B})$ as the vector containing the product between $\widehat{B}^T$, **T(R)** and $\widehat{B}$, which was the relevant quantity we used for μQSM in this study.

*Creating Masks for analysis*

Several masks required for subsequent analyses were derived from the co-registered images. A 3D binary tissue mask $M^{\text{Tissue}}(\mathbf{R})$ was constructed from the orientation-averaged $R_2$ maps, and its vectorized form denoted $\vec{M}^{\text{Tissue}} = \text{vec}(M^{\text{Tissue}}(\mathbf{R})) \in \{0,1\}^N$. From the MGE signal magnitudes of the first echo averaged over $\hat{\mathbf{B}}$ orientations, we generated a spherical mask $\vec{M}^{\text{Sph}}$ surrounding the tissue. The spherical mask was defined by computing the tissue center of mass and drawing a sphere centered at this point with a diameter 5 voxel layers (1.25 mm) smaller than that of the spherical container. This procedure helped mitigate image artifacts near the container surface. Lastly, we segmented the PBS by dilating the tissue mask by two voxel layers, referred to as $\vec{M}^{\text{Tissue}\prime}$, such that $\vec{M}^{\text{PBS}} = (\mathbf{I} - \vec{M}^{\text{Tissue}\prime}) \odot \vec{M}^{\text{Sph}}$, where $\odot$ defines point-wise mask (vector) multiplication The dilation ensured that we did not include the voxels closest layers to the tissue thereby avoiding partial volume effects.

**Analysis**

We now go through the two different data analyses performed to investigate the two objectives described in the introduction.

*O1) - Predicting the residual Larmor frequency shift in WM*

Similar to W&B[18], the experimental design enabled us to estimate tissue bulk magnetic susceptibility $\bar{\chi}^{\text{WM}}$ by looking at the induced frequency shift $\vec{f}_{\text{MRI}}$ in PBS only. The bulk susceptibility $\bar{\chi}^{\text{WM}}$ is assumed to be a single scalar value within the tissue. This simplification is justified given that the nerve is statistically homogeneous across its volume, allowing for a robust macroscopic estimation of susceptibility The idea behind this analysis is that, at macroscopic distances[7] (around 100 µm - 1 mm), the explicit microstructural organization of magnetized tissue becomes irrelevant. Instead, all we need for describing shifts at such long distances is the bulk susceptibility $\bar{\chi}^{\text{WM}}$ of the tissue inside the mask $\vec{M}^{\text{Tissue}}$. Consequently, frequency shifts measured inside the tissue, where anisotropic white-matter microstructure can generate additional sub-voxel contributions, were excluded from the fitting. In contrast, the frequency shift measured in PBS only, $\vec{f}_{\text{MRI}}$ restricted to $\vec{M}^{\text{PBS}}$, should be well described by the QSM forward model. The subsequent QSM-predicted shift inside the tissue is denoted $\vec{f}_{\text{QSM}}$. Hence, for all orientations $\hat{\mathbf{B}}$, the frequency shift $\vec{f}_{\text{QSM}}(\hat{\mathbf{B}})$ was modelled in QSM as

$$\begin{bmatrix} \vec{f}_{QSM}(\widehat{\mathbf{B}}_1) \\ \vec{f}_{QSM}(\widehat{\mathbf{B}}_2) \\ \vdots \\ \vec{f}_{QSM}(\widehat{\mathbf{B}}_n) \end{bmatrix} = \begin{bmatrix} \mathbf{A}^{QSM}(\widehat{\mathbf{B}}_1) \\ \mathbf{A}^{QSM}(\widehat{\mathbf{B}}_2) \\ \vdots \\ \mathbf{A}^{QSM}(\widehat{\mathbf{B}}_n) \end{bmatrix} \overrightarrow{\mathbf{M}}^{\text{Tissue}} \overline{\chi}^{WM} \qquad (1)$$

(QSM - Macroscopic frequency shift)

Notice $\overline{\chi}^{WM}$ is a scalar, i.e. independent of $\mathbf{R}$. Here $\mathbf{A}^{QSM}$ is an $N \times N$ matrix describes the dipole-field discrete convolution operation, with elements

$$[A^{QSM}(\widehat{\mathbf{B}})]_{i,j} = 2\pi\gamma B_0 \widehat{\mathbf{B}}^T \overline{\mathbf{Y}}(\mathbf{R}_i - \mathbf{R}_j) \widehat{\mathbf{B}}, \qquad (2)$$

which denotes the macroscopically induced frequency shift in voxel at position $\mathbf{R}_i$ from neighboring voxel at position $\mathbf{R}_j$, when the external field is at orientation $\widehat{\mathbf{B}}$. Similar to W&B, we applied background field removal to the simulated frequency shift $\vec{f}_{QSM}(\widehat{\mathbf{B}})$, an operation denoted as $F^{BFR}[\vec{f}_{QSM}]$. While no background fields are present in the simulation, applying $F^{BFR}$ to the simulated data ideally ensured that errors introduced by $F^{BFR}$ were the same for both simulation and measurements[18].

The minimization problem for estimating bulk magnetic susceptibility inside the container, using only $\vec{f}_{MRI}$ in PBS from all acquired orientations $\widehat{\mathbf{B}}$, can be written as

$$\overline{\chi}^{WM} = \underset{\chi'}{\text{argmin}} \sum_{\widehat{\mathbf{B}}} \left( \overrightarrow{\mathbf{M}}^{PBS} \odot \left( \mathbf{I} - \frac{1}{k_{PBS}} \overrightarrow{\mathbf{M}}^{PBS} (\overrightarrow{\mathbf{M}}^{PBS})^T \right) \right.$$
$$\left. F^{BFR}[\mathbf{A}^{QSM}(\widehat{\mathbf{B}}) \overrightarrow{\mathbf{M}}^{\text{Tissue}}] \chi' - \vec{f}_{MRI}(\widehat{\mathbf{B}}) \right)^2 \qquad (3)$$

The matrix $\left( \mathbf{I} - \frac{1}{n_{PBS}} \overrightarrow{\mathbf{M}}^{PBS} (\overrightarrow{\mathbf{M}}^{PBS})^T \right)$ implements referencing of the frequency shift to the PBS volume, analogous to the referencing applied to $\vec{f}_{MRI}$, where $n_{PBS}$ is the number of voxels in PBS. Before de-ironing (WFE), $\overline{\chi}^{WM}$ should reflect both myelin with susceptibility $\overline{\chi}^C$ and iron with susceptibility $\overline{\chi}^S$, such that $\overline{\chi}^{WM} = \overline{\chi}^C + \overline{\chi}^S$, and after de-ironing (DFE), $\overline{\chi}^{WM} = \overline{\chi}^C$ should reflect primarily myelinated axons sheaths including sources in the water bilayers[22].

*Residual frequency shift*

Using $\overline{\chi}^{WM}$ found with Eq. (3), we calculated $\vec{f}_{QSM}(\hat{\mathbf{B}})$ across the whole sample for each $\hat{\mathbf{B}}$ - including *inside* the tissue. From this we defined the residual frequency shift $\delta f(\hat{\mathbf{B}})$ across the whole sample as

$$\delta \vec{f}(\hat{\mathbf{B}}) \equiv \vec{f}_{MRI}(\hat{\mathbf{B}}) - \vec{f}_{QSM}(\hat{\mathbf{B}}), \qquad (4)$$

which captured every aspect that could not be accounted for with QSM. Here we postulate that $\delta f$, mainly stems from susceptibility-induced frequency shifts at the mesoscopic scale[21,22] rather than the macroscopic scale, and is therefore unaccounted for in conventional QSM. In the context of μQSM[21,22], the residual was then modelled as

$$\vec{f}^{Meso}(\hat{\mathbf{B}}) = -2\pi\gamma B_0 (\overline{\chi}^C - \lambda^S \overline{\chi}^S) \frac{1}{2} \left( \vec{T}(\hat{\mathbf{B}}) - \frac{1}{3} \right) + b_{other} \qquad (5)$$

inside the tissue. Recall that $\vec{T}(\hat{\mathbf{B}})$ was estimated from our dMRI experiment and governs the effect of orientation dispersion on the Larmor frequency shift. The scalars $\lambda^S \in (0,1)$ and $b_{other}$ are the only free parameters and control the contribution from anisotropically distributed magnetized spheres, and chemical shifts or other unmodelled frequency shifts, respectively. We expect $b_{other} \sim -1 Hz$ based on a previous experiment by Luo et al.[20,47]. The resulting tissue-averaged residual frequency shift, denoted as $\delta f$, was compared to the predicted mesoscopic frequency[21,22] using Eq. (5). For this we used the bulk susceptibilities $\overline{\chi}^C$ and $\overline{\chi}^S$, which were estimated from $\overline{\chi}^{WM}$ before and after de-ironing, and the tissue-averaged scatter matrix $T(\hat{\mathbf{B}})$. The susceptibility $\overline{\chi}^{WM}$ also provided a "ground truth" for later comparison in O2).

*O2) Susceptibility estimation from whole sample fitting using QSM and μQSM*

In O2), we performed a full fit across the whole container, including the tissue, to mimic conventional susceptibility fitting. We performed fitting using standard QSM, and μQSM assuming only cylindrical sources in Eq. (5). This choice was made to keep only a single degree of freedom in each voxel for both models and then compare the estimated susceptibility with $\overline{\chi}^{WM}$ from O1) - before and after de-ironing. The forward problem of QSM is

$$\begin{bmatrix} \vec{f}_{\text{QSM}}(\hat{\mathbf{B}}_1) \\ \vec{f}_{\text{QSM}}(\hat{\mathbf{B}}_2) \\ \vdots \\ \vec{f}_{\text{QSM}}(\hat{\mathbf{B}}_n) \end{bmatrix} = \begin{bmatrix} \mathbf{A}^{\text{QSM}}(\hat{\mathbf{B}}_1) \\ \mathbf{A}^{\text{QSM}}(\hat{\mathbf{B}}_2) \\ \vdots \\ \mathbf{A}^{\text{QSM}}(\hat{\mathbf{B}}_n) \end{bmatrix} \vec{\chi}^{\text{QSM}}, \qquad (6)$$

where in contrast to Eq. (1), we fitted a general position-dependent scalar susceptibility, vectorized as $\vec{\chi}^{\text{QSM}}$, to mimic conventional susceptibility fitting. In µQSM, the measured frequency shift was modelled as

$$\begin{bmatrix} \vec{f}_{\mu\text{QSM}}(\hat{\mathbf{B}}_1) \\ \vec{f}_{\mu\text{QSM}}(\hat{\mathbf{B}}_2) \\ \vdots \\ \vec{f}_{\mu\text{QSM}}(\hat{\mathbf{B}}_n) \end{bmatrix} = \begin{bmatrix} \mathbf{A}^{\text{QSM}}(\hat{\mathbf{B}}_1) + \mathbf{A}^{\text{C}}(\hat{\mathbf{B}}_1) \\ \mathbf{A}^{\text{QSM}}(\hat{\mathbf{B}}_2) + \mathbf{A}^{\text{C}}(\hat{\mathbf{B}}_2) \\ \vdots \\ \mathbf{A}^{\text{QSM}}(\hat{\mathbf{B}}_n) + \mathbf{A}^{\text{C}}(\hat{\mathbf{B}}_n) \end{bmatrix} \vec{\chi}^{\mu\text{QSM}}, \qquad (7)$$

where the added matrices describe the microstructure-related sub-voxel frequency shifts

$$[A^{\text{C}}(\hat{\mathbf{B}})]_{i,j} = -2\pi\gamma B_0 \frac{1}{2} \hat{\mathbf{B}}^{\text{T}} \left( \mathbf{T}(\mathbf{R}_j) - \frac{1}{3}\mathbf{I} \right) \hat{\mathbf{B}} M^{\text{Tissue}}(\mathbf{R}_j)\, \delta_{ij}. \qquad (8)$$

The inverse problems was be stated as

**QSM**:

$$\vec{\chi}^{\text{QSM}} = \underset{\vec{\chi}'}{\text{argmin}} \sum_{\hat{\mathbf{B}}} \left( \overrightarrow{M}^{\text{Sphere}} \odot \left( \mathbf{I} - \frac{1}{k_{\text{PBS}}} \overrightarrow{M}^{\text{PBS}} (\overrightarrow{M}^{\text{PBS}})^{\text{T}} \right) \right. \\
\left. \left( \mathbf{A}^{\text{QSM}}(\hat{\mathbf{B}}) \mathbf{M}^{\text{Sphere}} \vec{\chi}' - \vec{f}_{\text{MRI}}(\hat{\mathbf{B}}) \right) \right)^2 \qquad (9)$$

**µQSM:**

$$\vec{\chi}^{\mu\text{QSM}} = \underset{\vec{\chi}'}{\text{argmin}} \sum_{\hat{\mathbf{B}}} \left( \overrightarrow{M}^{\text{Sphere}} \odot \left( \mathbf{I} - \frac{1}{k_{\text{PBS}}} \overrightarrow{M}^{\text{PBS}} (\overrightarrow{M}^{\text{PBS}})^{\text{T}} \right) \right. \\
\left. \left( \left( \mathbf{A}^{\text{QSM}}(\hat{\mathbf{B}}) \mathbf{M}^{\text{Sphere}} + \mathbf{A}^{\text{C}}(\hat{\mathbf{B}}) \right) \vec{\chi}' - \vec{f}_{\text{MRI}}(\hat{\mathbf{B}}) \right) \right)^2. \qquad (10)$$

As in O1), all susceptibility values were subsequently referenced to PBS. After fitting, we compared the tissue-averaged $\overline{\chi}^{\text{QSM}}$ and $\overline{\chi}^{\mu\text{QSM}}$ to $\overline{\chi}^{\text{WM}}$, with (WFE) and without de-ironing (DFE). We also computed the average fitting residual $\delta f_{\mu\text{QSM}}(\hat{\mathbf{B}}) = f_{\text{MRI}}(\hat{\mathbf{B}}) - f_{\mu\text{QSM}}(\hat{\mathbf{B}})$ and $\delta f_{\text{QSM}}(\hat{\mathbf{B}}) = f_{\text{MRI}}(\hat{\mathbf{B}}) - f_{\text{QSM}}(\hat{\mathbf{B}})$. Here the residual should be close to zero, and invariant of $\hat{\mathbf{B}}$.

*Transverse relaxation rate*

Lastly, we estimated the effective transverse relaxation rate $R_2^\star$ across two ranges of echo times, $T_E = 7,19,30$ ms or $T_E = 43,55,67$ ms. This was done by fitting the signal magnitude to an exponentially decaying function for each sample orientation, with and without de-ironing. The rate was averaged across the tissue for each orientation.

# 4 | Results

*fODF*

From dMRI analysis, we successfully estimated the scatter matrix $\vec{\mathbf{T}}(\mathbf{R})$. Analysis revealed that the orientation dispersion angle $\theta_{P2}$ was around 25 degrees before de-ironing and 29 degrees after. A similar result was found for a PFA-fixed optic nerve set (27 degrees versus 30 degrees) - where the same nerve was scanned and analyzed before and after de-ironing (see supplementary material). This may indicate that de-ironing influenced the axonal microstructure (*see also results on transverse relaxation below*).

*O1) - Predicting the residual Larmor frequency shift in WM*

Figure 4 shows the measured frequency shift $\vec{f}_{MRI}$ after phase unwrapping and background-field removal using LBV, the simulated frequency shift $\vec{f}_{QSM}$, and lastly the residual $\delta\vec{f}$ between the two. A clear positive bias was visible in the tissue, when the axons were colinear with the field, while a negative bias was present when they were perpendicular. The residual outside the tissue did not exhibit any systematic behavior with angle, but fluctuations appeared due to image distortions and ringing in the actual data, which were caused by bubbles on the tissue and spherical container surface.

Table 1 shows the estimated susceptibility before (-146 ppb) and after de-ironing (-159 ppb). Notice that the susceptibility values were more negative compared to what is reported for typical fixated WM[18,20]. This is not unexpected due to GA in the fixation protocol[34]. De-ironing only induces a very small negative change in susceptibility, which may be indicative of very little iron being removed. We also analyzed another optic nerve sample fixed using only PFA, and there we found (see supplementary material) susceptibility around -82.9 ppb before de-ironing, which agreed with previous findings[20], and -74.8 ppb after de-ironing. Here the change in susceptibility was positive - opposite to the GA protocol and to what should be expected if iron with positive susceptibility was removed. Hence, we suspect that optic nerve did not contain a substantial amount of iron, or that it was already diluted during fixation[48].

*Residual frequency shift $\delta f$*

Figure 5 shows the average residual Larmor frequency shift $\delta f$ inside the tissue for different orientations $\angle(\boldsymbol{t}, \hat{\mathbf{B}})$ between the nerve and $B_0$. Here it was clear that the angle-dependent residual shift follows a $\sin^2$ dependence with the angle between axons and external field $\mathbf{B_0}$. The reisudal residual $\delta f$ was then compared to $f^{Meso}$. For this we used the dMRI-estimated scatter matrix $T(\hat{\mathbf{B}})$, the estimated magnetic susceptibilities before and after de-ironing, and fitted the parameters $\lambda^S$ and $b_{other}$ to minimize the difference between $\delta f$ and $f^{Meso}$. As listed in Table 1, $\lambda^S = 0$, indicating no impact from anisotropically positioned iron in the nerve. A small offset of less than -1 Hz produced the best agreement between $\delta f$ with $f^{Meso}$ before de-ironing, while no offset was found after de-ironing. This means that Eq. (5) predicted $\delta f$, and that only cylindrical susceptibility sources were needed to explain the residual shift measured here.

*O2) Susceptibility estimation from whole sample fitting using QSM and µQSM*

Figure 5 shows the fitted susceptibility of the whole sample using either QSM or µQSM, including the difference between the two images. A clear negative difference in susceptibility was found inside the tissue. As listed in Table 1, QSM estimated susceptibility to be -112.6 ppb and -127.7 ppb before and after de-ironing, respectively. µQSM yielded a susceptibility of -146.9 ppb and -159.1 ppb before and after de-ironing. Compared to -146.2 ppb and -158.8 ppb found in O1), we found that QSM had an deviated around 20-22%, while µQSM deviated around 1%. Similar estimation error of 13-18% for QSM and 1-2% for µQSM was also found for the pure PFA-fixed nerve. The QSM error agreed with previous estimated estimation errors in ex vivo mouse brain[21]. Figure 7 shows the residual Larmor frequency shift $\delta f$. In contrast to O1) we wanted the residual for µQSM to be zero and exhibit no orientation dependence. Here we found that QSM was clearly biased depending on the field direction, while µQSM showed a lower bias and no distinguishable orientation dependence.

*Transverse relaxation*

Figure 8 shows the estimated effective relaxation rate. Before de-ironing, the relaxation's orientation dependence was close to $\sin^4$ and decreased when estimated at longer echo times. This decrease was expected, and likely due to fast relaxation of myelin water contributing less to the signal for the long echo times. Remarkably, after de-ironing, we found $R_2^\star$ increased compared to before de-ironing. If iron *were* removed, we expected the relaxation rate to go down. Instead, the increase appeared mainly to be driven by increased orientation anisotropy, and effect resembling what was expected from increased axonal beading[49] and increased vacuolization[50]. In supplementary material, we show that the relaxation rate of the optic nerve fixed in pure PFA, did not show any substantial change in $R_2^\star$ magnitude before versus after de-ironing. Contrary to the GA+PFA fixed nerve set, for pure PFA fixation, de-ironing changed the orientation dependence such that it resembled a $\sin^4$ dependence. This varying behavior may be indicative of different microstructural changes induced depending on the fixation and de-ironing protocol used.

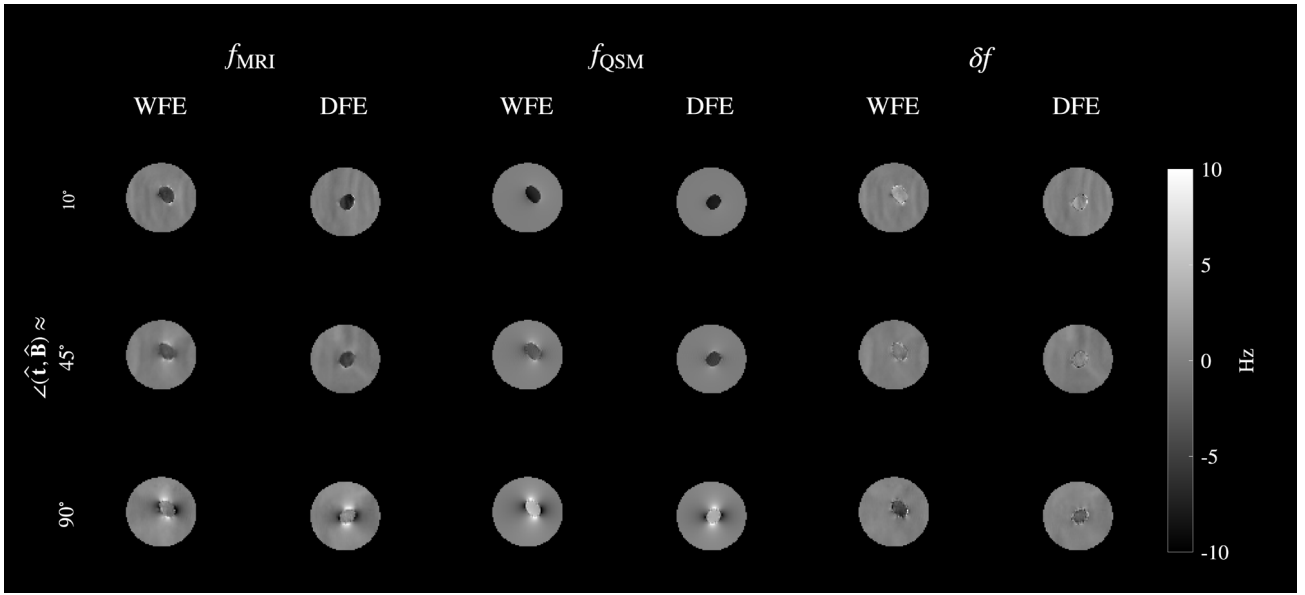

Figure 4 - Larmor frequency shift $f_{\text{MRI}}$ inside and outside the optic nerve are shown the first columns. The third and fourth column shows the simulated frequency $f_{\text{MRI}}$, while the fifth and sixth shows the residual $\delta f$. The frequency shifts are shown in a slice perpendicular to the main direction of the nerve. Each column set corresponds to before (WFE) and after (DFE) de-ironing. Rows show the frequency shift for different B0 orientations relative to $\hat{t}$, the average orientation of the major eigenvector of the scatter matrix **T** estimated from dMRI.

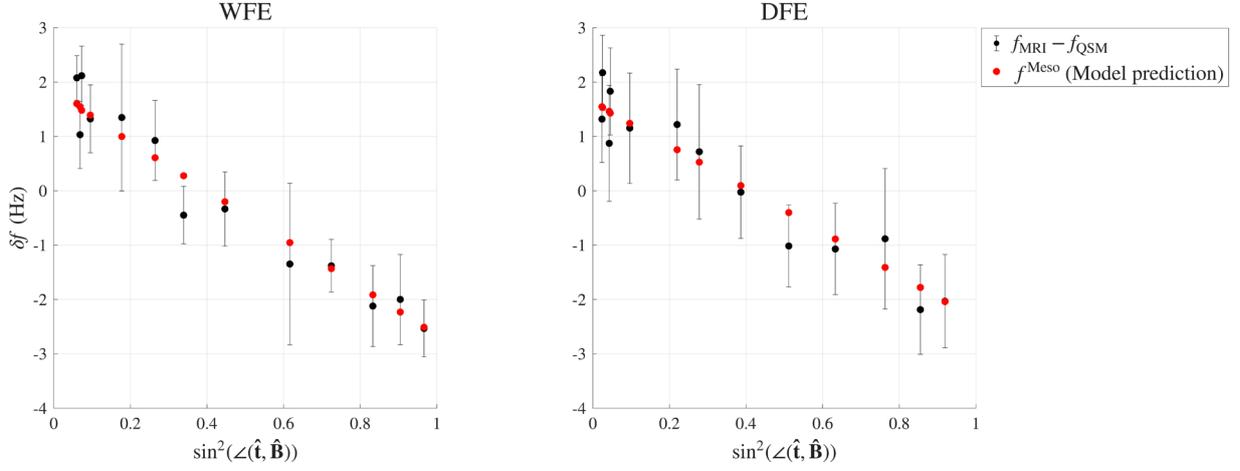

Figure 5 – Tissue average residual Larmor frequency shift $\delta f$ inside the optic nerves. The left graph shows results with iron (WFE), and right one after de-ironing (DFE). The *x*-axis denotes different angles between the nerve and $B_0$ field. Here $\hat{t}$ denotes the average orientation of the major eigenvector of the scatter matrix **T** estimated from dMRI. Black points show mean residual $\delta f$ estimated from the measured shift $f_{MRI}$ and $f_{QSM}$ estimated from fitting *only* the frequency values in PBS. Red points show the µQSM-predicted frequency shift $f^{Meso}$ using Eq. (5). The black error bars show the standard deviation of $\delta f$, while the red error bars of $f^{Meso}$ are derived using the 95% confidence intervals of the fitting parameters.

| **O1)** | | | | | |
|---|---|---|---|---|---|
| | $\overline{\chi}^{WM}$ | $\overline{\chi}^{C}$ | $\overline{\chi}^{S}$ | $\lambda^{S}$ | $b_{other}$ |
| WFE | $-146.2 \pm 0.3$ ppb | $-159$ ppb | $13$ ppb | $0.0 \pm 1$ | $-1.2 \pm 0.3$ Hz |
| DFE | $-158.8 \pm 0.3$ ppb | | | | $-0.9 \pm 0.2$ Hz |
| **O2)** | | | | | |
| | $\overline{\chi}^{QSM}$ | | | $\overline{\chi}^{\mu QSM}$ | |
| WFE | $-112.6$ ppb | | | $-146.9$ ppb | |
| DFE | $-127.7$ ppb | | | $-159.1$ ppb | |

*Table 1 – Susceptibility values $\overline{\chi}^{WM}$ estimated by fitting the Larmor frequency in PBS, induced by WM tissue – with (WFE) and without (DFE) iron. The myelin and iron related susceptibility $\overline{\chi}^C$ and $\overline{\chi}^S$ were then estimated by appropriate combinations of the two. $\lambda^S$ denotes the weight of the spherical contribution, while $b_{other}$ is the estimated frequency shift in tissue not accounted for by our model. 95% confidence from parameter fitting are listed for **O1**).*

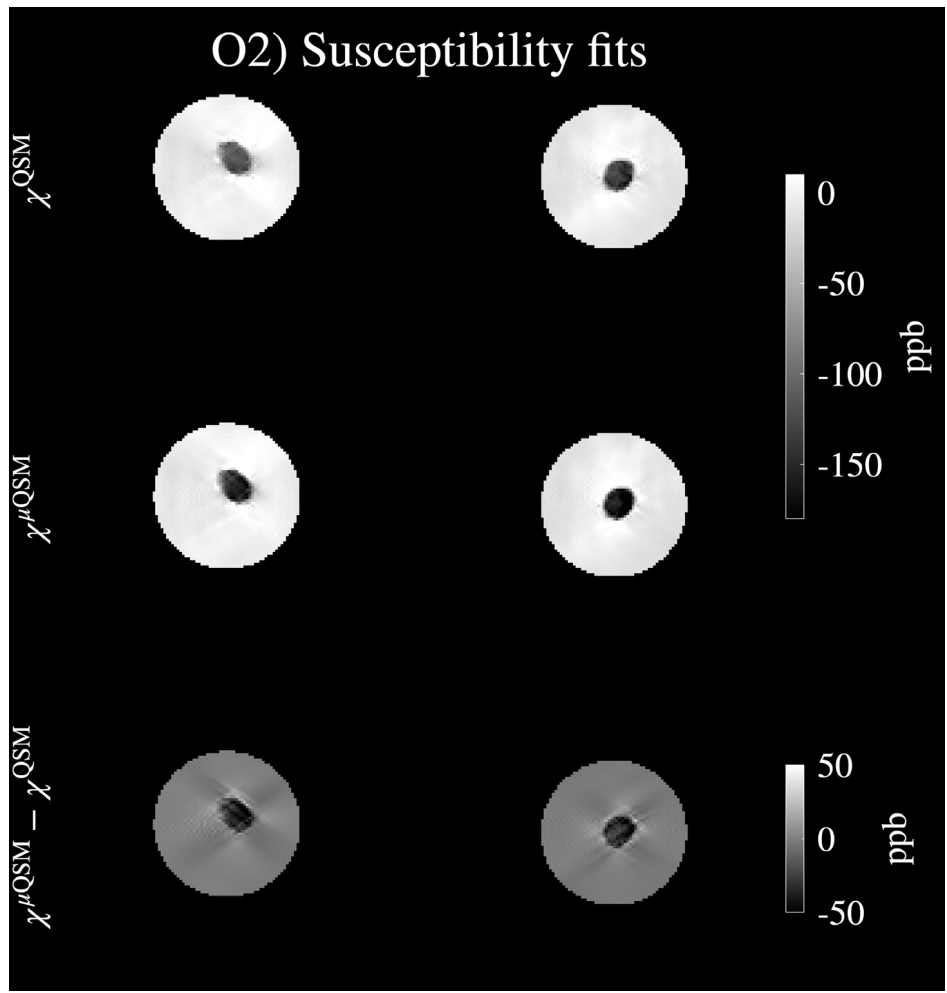

Figure 6 – The first row shows susceptibility $\vec{\chi}^{QSM}$ in the optic nerve fitted using QSM, the second row shows $\vec{\chi}^{\mu QSM}$ estimated by μQSM, while the third row shows the difference between the two.

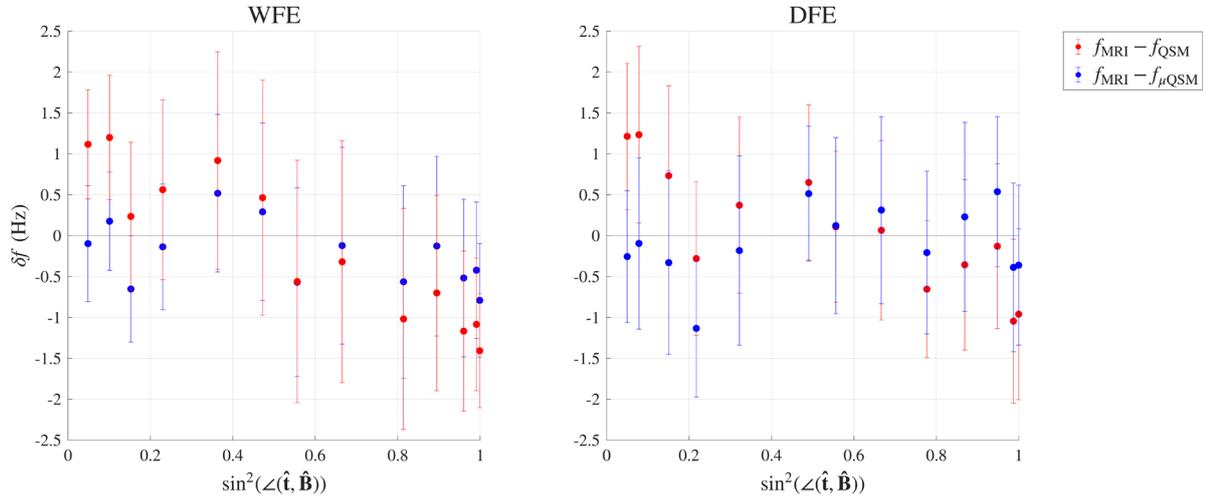

Figure 7 - Tissue average residual Larmor frequency shift $\delta f$ inside the optic nerves. Red points show residual $\delta f$ estimated from the measured shift $f_{MRI}$ and $f_{QSM}$ estimated from fitting *all* the frequency values in the sample. Blue points show fitting residual $\delta f$ estimated from the measured shift $f_{MRI}$ and $f_{\mu QSM}$ using µQSM. The left graph shows with iron (WFE), and right graph after de-ironing (DFE). The *x*-axis denotes different angles between the nerve and $B_0$ field. Here $\hat{t}$ denotes the average orientation of the major eigenvector of the scatter matrix **T** estimated from dMRI. The error bars show the standard deviation of $\delta f$.

Transverse Relaxation

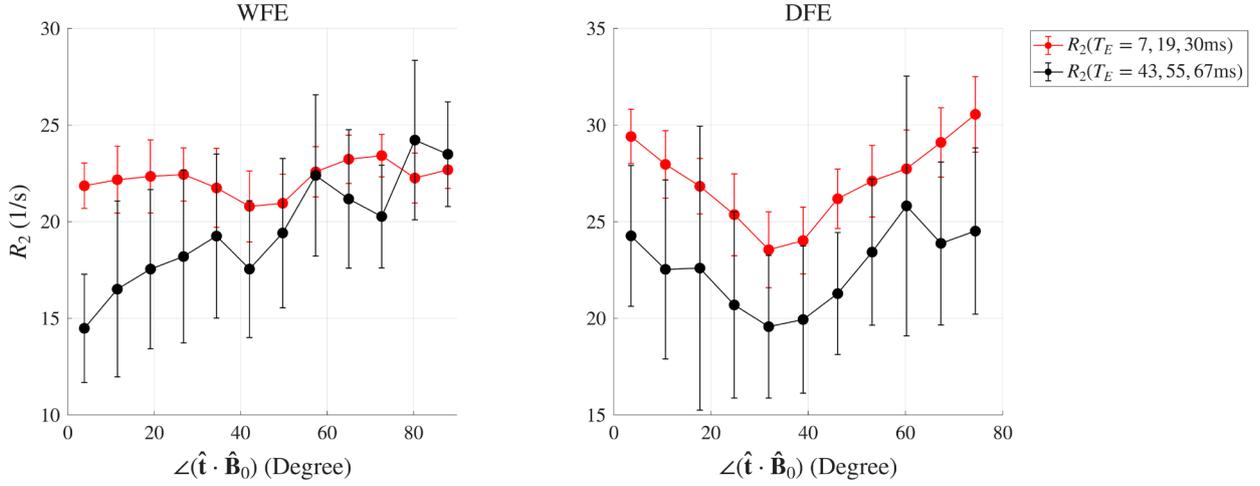

Figure 8 – Tissue-averages of transverse relaxation rate $R_2^\star$ for different $B_0$ orientations. Left graph shows before de-ironing, while the right graph shows after. $R_2^\star$ is fitted using three echo times, either between 7-30 ms (red) or 43-67 ms (black). The error bars show the standard deviation of $R_2^*$

# 5 | Discussion

In this study, we aimed to validate μQSM - an extension of QSM[9] that incorporates microstructure-related mesoscopic frequency shifts in white matter - to achieve quantitative estimates of magnetic susceptibility. By reproducing the experimental paradigm of W&B[18], we confirmed the presence of a residual "sub-voxel" frequency shift in WM that conventional QSM cannot account for. Neglecting such a microstructural frequency shift led to a substantial bias around 20% for the estimated bulk magnetic susceptibility. Crucially, by combining dMRI and susceptibility data[21,23], we demonstrated that this residual shift could be **predicted** by the μQSM model[7,21,22], which treats the tissue as a collection of structurally anisotropic, orientationally dispersed axons with scalar susceptibility. Diffusion MRI, therefore, plays a crucial role in estimating mesoscopic effects of anisotropic tissue in susceptibility estimation. The presented results here also support our recent Monte-Carlo simulations[27].

We therefore recommend using μQSM for predicting magnetic susceptibility in WM structures like the optic nerve, for example corpus callosum or optic radiation. We aim to examine more brain WM tissue structures in future experiments to confirm the general validity of μQSM.

**Limitations**

*Effects of fixation on diamagnetic susceptibility*

Our results highlight that while the choice of fixative influences the magnetic properties of the tissue, these variations do not violate the underlying model assumptions in μQSM. Instead, the chemical effects of fixation are directly reflected in the model's estimated parameters. For example, we observed that nerves fixed with a PFA/GA combination exhibited more negative (diamagnetic) susceptibility than those fixed with PFA alone. This is consistent with the 'additive' nature of GA, where permanent molecular integration and increased crosslinking density alter the local electronic environment[51,52]. While these structural and electronic shifts enhance the bulk diamagnetic response[47], the framework remained internally consistent, and captured these systematic differences as distinct parameter values rather than modeling errors. Hence, while susceptibility values depend on preparation, the model provides a reliable quantitative characterization of the tissue in its specific prepared state.

*Effects of de-ironing*

The contribution of iron to the susceptibility of the porcine optic nerve appeared remarkably low. De-ironing produced only a minor shift in susceptibility of approximately 10 ppb, resulting in a negligible effect on the sub-voxel frequency shift. A reason for this low effect of de-ironing may be the previous effect of the fixation process on transition metals. As reported by Schrag et al.[48], iron concentration in fixed tissue can be reduced by approximately 43% on average due to leaching into the fixative solution during storage. This suggests that the "baseline" iron content in our samples may already have been depleted before our experimental de-ironing protocol began. Furthermore, we observe unexpected changes in tissue properties following the de-ironing process, specifically an increase in $R_2^*$, its anisotropy and in the dMRI estimated orientation dispersion. Typically, removing iron should reduce the transverse relaxation rate. Hence, the fact that it both increased and decreased depending on fixation protocol suggests that the tissue structure was influenced as well. Chelation of metals may therefore have compromised the structural integrity of the axonal membranes, if the chelator, for example, removed divalent ions essential for stabilizing the membrane and intra-axonal skeleton. Such effect could mimic axonal beading or induce vacuolization, and increase the orientation dependence[49] in $R_2^\star$. We emphasize that our proposed explanation remains speculative, as answering this question is beyond the focus of this study. Nevertheless, we aim to consider the effects of fixation and de-ironing in future studies, where we will include electron-microscopy and iron quantification. We also plan to investigate the behavior of transverse relaxation more systematically in the future. In other parts of the brain, such as superficial WM, iron may inarguably play a greater role[11,12]. Here we suspect that sub-voxel frequency shifts from iron may be non-negligible. We thus plan to perform a similar experiment and analysis of more iron-rich tissue in the future.

*Limitations in experimental setup*

Several practical limitations related to our experimental setup and data acquisition influenced the precision of our measurements. The implementation of a bespoke rotatable RF coil setup was essential for sampling the tissue at multiple orientations with high resolution at 3T. However, our images still suffered from localized image distortions and phase artifacts that required careful consideration during registration. Furthermore, while the isotropic resolution of 250 μm in this study was sufficient for characterizing the porcine optic nerve, future experiments would benefit from even higher spatial resolutions - ideally 125 μm or finer - or a larger nerve. Such an increase in resolution would enable even better tissue segmentation and further minimize partial volume effects, which are particularly relevant at the boundaries of the nerve and surrounding medium. Such partial volume effects and image distortions can affect the estimated susceptibility value in O1) and images in O2). We emphasize, while such high resolution would be beneficial in validation studies, it is not a prerequisite for using μQSM in a clinical or pre-clinical setting. Another challenge was the presence of air bubbles on the tissue surface, which were remarkably difficult to eliminate - even though PBS, as used here, should reduce the problem compared to agar used by W&B. Despite rigorous protocols involving shaking of the sample and the application of vacuum degassing, small air pockets, or impurities, remained trapped on the tissue surface. These bubbles can be problematic because they create localized magnetic field perturbations that can affect how well we can estimate the mesoscopic frequency shifts. Finally, it can be mentioned as a limitation that the study was conducted on pigs that had been anesthetized with ketamine-midazolam-propofol during a previous PET scan experiment and subsequently euthanized with pentobarbital. Pentobarbital is the standard for euthanasia of pigs for experiments but is known to have some minor histological effects on the tissue[53].

In future studies, the use of a larger sample sphere might provide a more uniform field environment and help distance the tissue from surface-related artifacts; however, this approach involves a significant trade-off, as a larger sphere necessitates the use of larger RF-coils, which inherently results in a lower SNR. Optimizing the experimental setup will also be pursued in future studies, aiming to test µQSM across multiple field strengths, and tissue types other than optic nerve.

## 6 | Conclusion

This study demonstrates that microstructure-informed quantitative susceptibility mapping (µQSM) provides a robust and accurate framework for predicting mesoscopic Larmor frequency shifts in white matter. By validating our model against experimental data from porcine optic nerves, we have shown that incorporating axonal microstructure, characterized by the orientation distribution function estimated with diffusion MRI, effectively eliminates the orientation-dependent bias that hampers conventional QSM, and can provide more faithful susceptibility values. Our findings reveal that fixation and de-ironing can lead to susceptibility shifts and potential structural damage during de-ironing, which highlights the need for careful sample preparation in ex vivo studies. Ultimately, our work shows that the integration of microstructural information into the QSM model is a vital step toward achieving faithful, quantitative imaging of magnetic susceptibility and we recommend the adoption of µQSM for future investigations of white matter magnetic susceptibility.

## 7 | Funding

This study is funded by the Independent Research Fund (grant number 10.46540/3103-00144B).

## 8 | Data availability

Data is available upon reasonable request.

## 9 | References

# Supplementary Materials

# Predicting Mesoscopic Larmor Frequency Shifts in Ex Vivo Porcine Optic Nerve

*Anders Dyhr Sandgaard[1,2], André Pampel[2], Roland Müller[2], Niklas Wallstein[2,3], Toralf Mildner[2], Carsten Jäger[4,5], Markus Morawski[5], Aage Kristian Olsen Alstrup[6,7], Harald E. Möller[2,8], Sune Nørhøj Jespersen[1,9]*

[1]*Center of Functionally Integrative Neuroscience, Department of Clinical Medicine, Aarhus University, Aarhus, Denmark,*

[2]*NMR Methods & Development Group, Max Planck Institute for Human Cognitive and Brain Sciences, Leipzig, Germany*

[3]*Aix Marseille Univ, CNRS, CRMBM, Marseille, France*

[4]*Department of Neurophysics, Max Planck Institute for Human Cognitive and Brain Sciences, Leipzig, Germany,*

[5] *Paul Flechsig Institute - Center of Neuropathology and Brain Research, Medical Faculty, University of Leipzig, Leipzig, Germany,*

[6]*Department of Clinical Medicine, Aarhus University, Aarhus, Denmark*

[7]*Department of Nuclear Medicine, Aarhus University Hospital, Aarhus, Denmark.*

[8]*Felix Bloch Institute for Solid State Physics, Leipzig University, Leipzig, Germany*

[9]*Department of Physics and Astronomy, Aarhus University, Aarhus, Denmark*

**Sample Preparation of Set #2**

The second set consisted of a single ON, instead of a pair like set #2. Immediately after extraction, the nerve was immersed in PBS containing 4% PFA and stored in a refrigerator for one week. It was then washed in PBS and the collagen-rich epineurium was removed. The nerve was mounted in a similar spherical container and filled with PBS (cf. Figure 2 in manuscript). Here we used copper wire for mounting as suggested by (Luo et al., 2014). The nerve was then imaged with MRI (see description below). After imaging, the PBS *inside* the sphere was replaced with de-ironing solution as described by (Brammerloh et al., 2021), which was refreshed regularly over 12 days until clear. This means that the nerve was kept in the sphere, during de-ironing. This allowed imaging of the same nerve before and after iron removal without repositioning. After de-ironing, the sphere was refilled with PBS and the nerve was re-imaged.

**Imaging**

Imaging was done in a similar way to the protocol described in the manuscript, but with two differences: MGE echo times ranged from 18.65 ms to 53.25 ms in steps of 8.65 ms for set #2, and bipolar readout gradients were used.

**Results**

Figure S1 shows the residual frequency shift, and the predicted shift using Eq. (5) in the manuscript. Analysis was done similar to O1) in manuscript.

Figure S2 show the transverse relaxation rate.

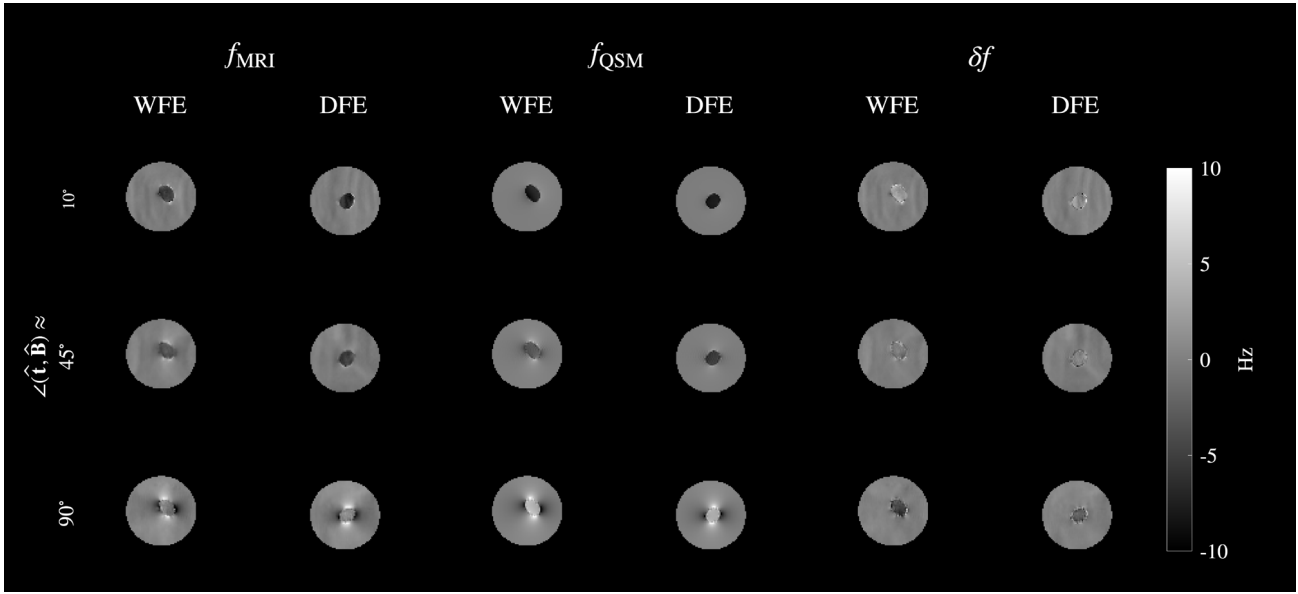

Figure S1 - Larmor frequency shift $f_{MRI}$ inside and outside the optic nerve fixed with PFA only are shown the first columns. The third and fourth column shows the simulated frequency $f_{MRI}$, while the fifth and sixth shows the residual $\delta f$. The frequency shifts are shown in a slice perpendicular to the main direction of the nerve. Each column set corresponds before (WFE) and after (DFE) de-ironing. Rows show the frequency shifts for different B0 orientations relative to $\hat{t}$, the average orientation of the major eigenvector of the scatter matrix **T** estimated from dMRI.

Figure S2 – Tissue average residual Larmor frequency shift $\delta f$ inside the optic nerves. Column one shows with iron (WFE), and second column after de-ironing (DFE). The *x*-axis denotes different angles between the nerve and $B_0$ field. Here $\hat{t}$ denotes the average orientation of the major eigenvector of the scatter matrix **T** estimated from dMRI. Black points show residual $\delta f$ estimated from the measured shift $f_{\text{MRI}}$ and $f_{\text{QSM}}$ estimated from fitting *only* the frequency values in PBS. Red points show the μQSM-predicted frequency shift $f^{Meso}$ using Eq. (5).

Figure S3 - First row shows susceptibility $\vec{\chi}^{\text{QSM}}$ of optic nerve fit using QSM, the second row shows $\vec{\chi}^{\mu\text{QSM}}$ from μQSM, while the third row shows the difference between the two.

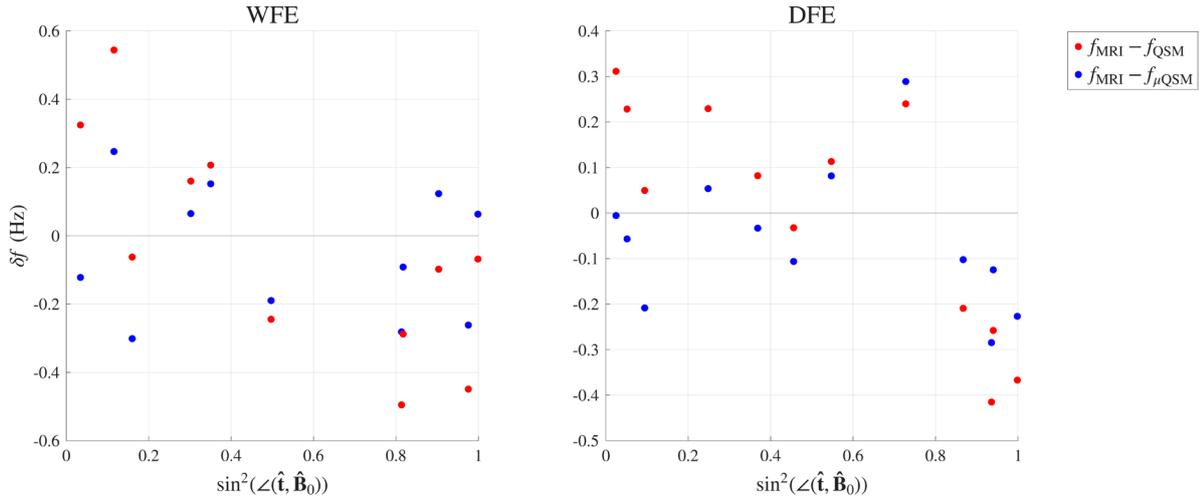

Figure S4 - Tissue average residual Larmor frequency shift $\delta f$ inside the optic nerves. Red points show residual $\delta f$ estimated from the measured shift $f_{MRI}$ and $f_{QSM}$ estimated from fitting *all* the frequency values in the sample. Blue points show residual $\delta f$ estimated from the measured shift $f_{MRI}$ and $f_{\mu QSM}$ using µQSM. Column one shows data before (WFE), and second column after de-ironing (DFE). The *x*-axis denotes different angles between the nerve and $B_0$ field. Here $\hat{t}$ denotes the average orientation of the major eigenvector of the scatter matrix **T** estimated from dMRI.

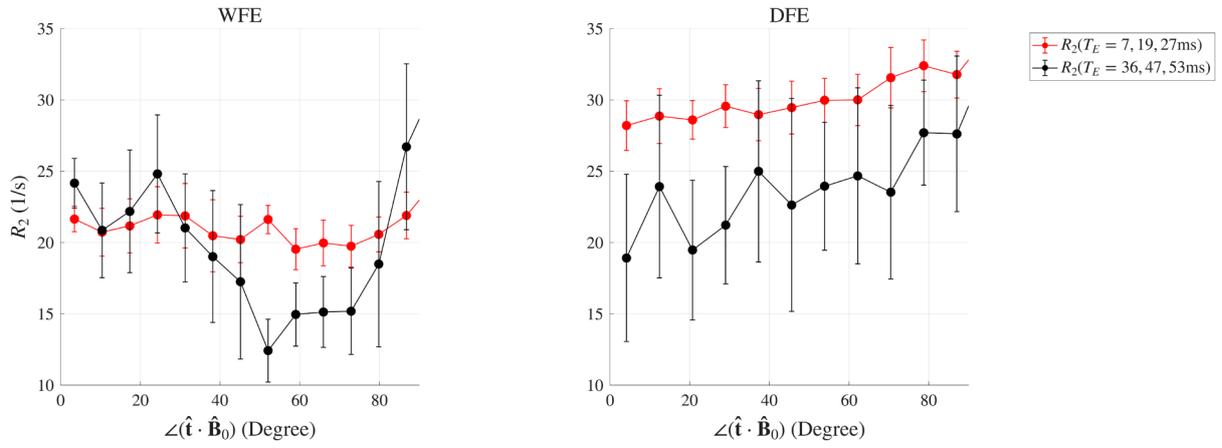

Figure S5 - Tissue-averaged transverse relaxation $R_2$ for different $B_0$ orientations. Left image shows before de-ironing, while the right shows after. $R_2$ is fitted using three echo times, either between 7-27 ms (red) or 36-53 ms (black).